\documentclass[10pt,showpacs,amsmath,amssymb,floatfix,superscriptaddress,longbibliography,twocolumn,eqsecnum]{revtex4-2}
\usepackage{booktabs}
\usepackage{amsmath}
\usepackage{physics}
\usepackage{graphicx}
\usepackage[usenames,dvipsnames,svgnames,table]{xcolor}
\usepackage{tcolorbox}
\counterwithout{equation}{section}
\counterwithout{figure}{section}
\usepackage[unicode=true,
            pdfusetitle,
            bookmarks=true,
            bookmarksnumbered=false,
            bookmarksopen=false,
            breaklinks=true,
            pdfborder={0 0 0},
            backref=false,
            colorlinks=true,
            hypertexnames=false]{hyperref}
\hypersetup{linkcolor=NavyBlue,urlcolor=NavyBlue,citecolor=NavyBlue}

\begin{document}

\title{Recognition of Schrödinger cat state based on CNN}
\author{Tao Zhang}   
\affiliation{School of Sciences, Hangzhou Dianzi University, Hangzhou 310018, China}
\author{Chaoying Zhao}
\email{zchy49@163.com}
\affiliation{School of Sciences, Hangzhou Dianzi University, Hangzhou 310018, China}
\affiliation{State Key Laboratory of Quantum Optics and Quantum Optics Devices, Institute of Opto-Electronics, Shanxi University, Taiyuan 030006, China}
\affiliation{Zhejiang Key Laboratory of Quantum State Control and Optical Field Manipulation, Hangzhou Dianzi University, Hangzhou 310018, China}

\begin{abstract}

We applied convolutional neural networks to the classification of cat states and coherent states. Initially, we generated datasets of Schrödinger cat states and coherent states from nonlinear processes and preprocessed these datasets. Subsequently, we constructed both LeNet and ResNet network architectures, adjusting parameters such as convolution kernels and strides to optimal values. We then trained both LeNet and ResNet on the training sets. The loss function values indicated that ResNet performs better in classifying cat states and coherent states. Finally, we evaluated the trained models on the test sets, achieving an accuracy of 97.5$\%$ for LeNet and 100$\%$ for ResNet. We evaluated cat states and coherent states with different $\alpha$, demonstrating a certain degree of generalization capability. The results show that LeNet may mistakenly recognize coherent states as cat states without coherent features, while ResNet provides a feasible solution to the problem of mistakenly recognizing cat states and coherent states by traditional neural networks.

\end{abstract}

\maketitle

%\begin{quotation}

%\end{quotation}

\section{\label{sec:level1}INTRODUCTION}

Schrödinger's cat state \cite{monroe1996schrodinger,leibfried2005creation} is a far-reaching concept in quantum mechanics, which was proposed by physicist Ernest Schrödinger in 1935. This concept aims to show the peculiar properties of quantum superposition states, and explore the measurement and observation problems in quantum mechanics by imagining a macro scale quantum superposition state -- a cat in a state of "life" and "death". Although this assumption is difficult to achieve in practical experiments, its theoretical value and the discussion of the basic problems of quantum mechanics make the Schrödinger cat state of great significance in the fields of quantum information processing \cite{vlastakis2013deterministically,ourjoumtsev2006generating}, quantum computing \cite{mirrahimi2016cat} and quantum communication.

In quantum information science, Schrödinger cat state is usually used as a key tool for testing and verifying quantum computing and quantum communication systems. They can help researchers evaluate the coherence of quantum states, measurement accuracy and stability of quantum systems. However, the accurate identification and classification of Schrödinger cat states still face many challenges in experimental and theoretical research. Traditional measurement and analysis methods may be cumbersome and inflexible when dealing with high-dimensional quantum states, so new technical means are urgently needed to improve efficiency and accuracy.

In recent years, the development of deep learning, especially convolutional neural network \cite{li2021survey,gu2018recent,o2015introduction,yamashita2018convolutional,aghdam2017guide} (CNN), has provided new possibilities for solving this problem. CNN is a neural network model that automatically extracts and learns image features by imitating biological vision system. Unlike traditional feature extraction methods, CNN can automatically learn complex features and patterns from data, and show excellent performance in image classification, target detection and other tasks. Due to the high-dimensional data and complexity of quantum states, CNN's feature learning ability provides a powerful tool for the recognition of quantum states.

In the field of quantum state recognition, especially for complex quantum states such as Schrödinger cat state, the application of CNN has important research value and potential \cite{schmale2022efficient,harney2020entanglement}. Wigner function is a tool commonly used to describe quantum states. It represents the density matrix of quantum states as a two-dimensional distribution image of coherent states. By analyzing the image of Wigner distribution, researchers can obtain rich information of quantum states. However, the complexity and high-dimensional features of Wigner images make it difficult for traditional image processing methods to classify and recognize effectively \cite{sinha2014scene,cai2019classification,zaman2021classification}.

Therefore, the application of CNN to the recognition of Schrödinger cat states can automatically extract key features from Wigner images by using its powerful image analysis and pattern recognition capabilities, so as to improve the classification accuracy and efficiency of Schrödinger cat states. This will not only help deepen our understanding of Schrödinger cat state, but also promote other related applications in quantum information science.

The purpose of this paper is to explore the Schrödinger cat state recognition method based on convolutional neural network. We will introduce how to apply CNN model to the recognition of quantum states, especially to classify Schrödinger cat states and coherent states by analyzing Wigner distribution images. First, we will describe the basic principle of CNN and its application in image classification, and then elaborate how to combine CNN with quantum state recognition task, and propose the corresponding model design and training methods. Then, we will show the experimental results, evaluate the performance of residual network (ResNet) in Schrödinger cat state recognition, and compare it with the traditional method (LeNet).

\section{Dataset}
The dataset we use is composed of the multi-component Schrödinger cat state generated by the nonlinear Hamiltonian \cite{he2023fast} and the most primitive coherent state. These quantum states are prepared into a two-dimensional Wigner distribution map by Wigner function. The original image resolution is 512*512 and the number of channels is 4. In addition to the RGB three channels, it also has a parameter $\alpha$ to measure the transparency of the image.

We output 100 pictures of coherent state, 2-body cat state, 3-body cat state and 4-body cat state respectively, corresponding to the photon number from 1 to 100, so we can get the distribution map of different forms. When n=56, the coherent characteristics of Schrödinger cat state become very indistinct after superposition, and it is difficult to distinguish cat state and coherent state according to the naked eye.

Perform data preprocessing on 400 pictures. First, adjust the resolution of the pictures to 128*128, then convert the data type to tensor, set the batch size to 16, define the data set, and put the picture data into the data set. We randomly select 6 pictures from a batch for display, as shown in Figure 1.

\begin{figure}
\includegraphics[width=0.48\textwidth]{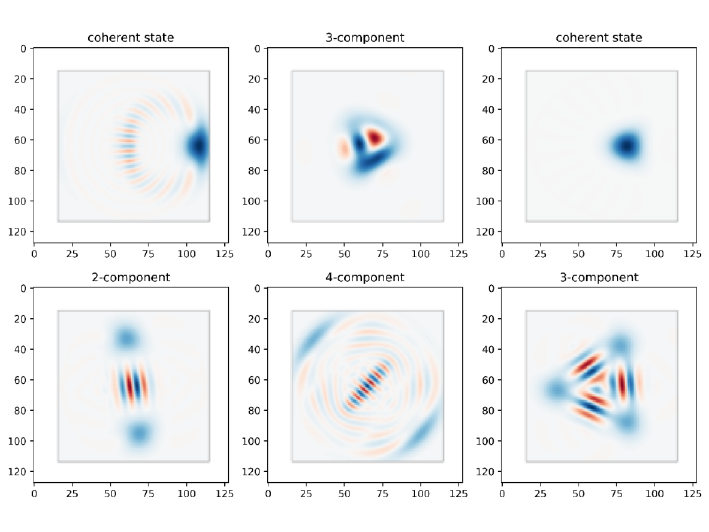}% Here is how to import EPS art
\caption{\label{fig:1} Preprocessed dataset. Randomly select 6 pictures of cat state and coherent state from a batch for display.}
\end{figure}

We selected 80$\%$ of the dataset as the training set, and the remaining 20$\%$ as the test set, and randomly scrambled the images in the data set to prevent over fitting.

\section{theoretical model}

\subsection{LeNet}

\begin{figure}
\includegraphics[width=0.48\textwidth]{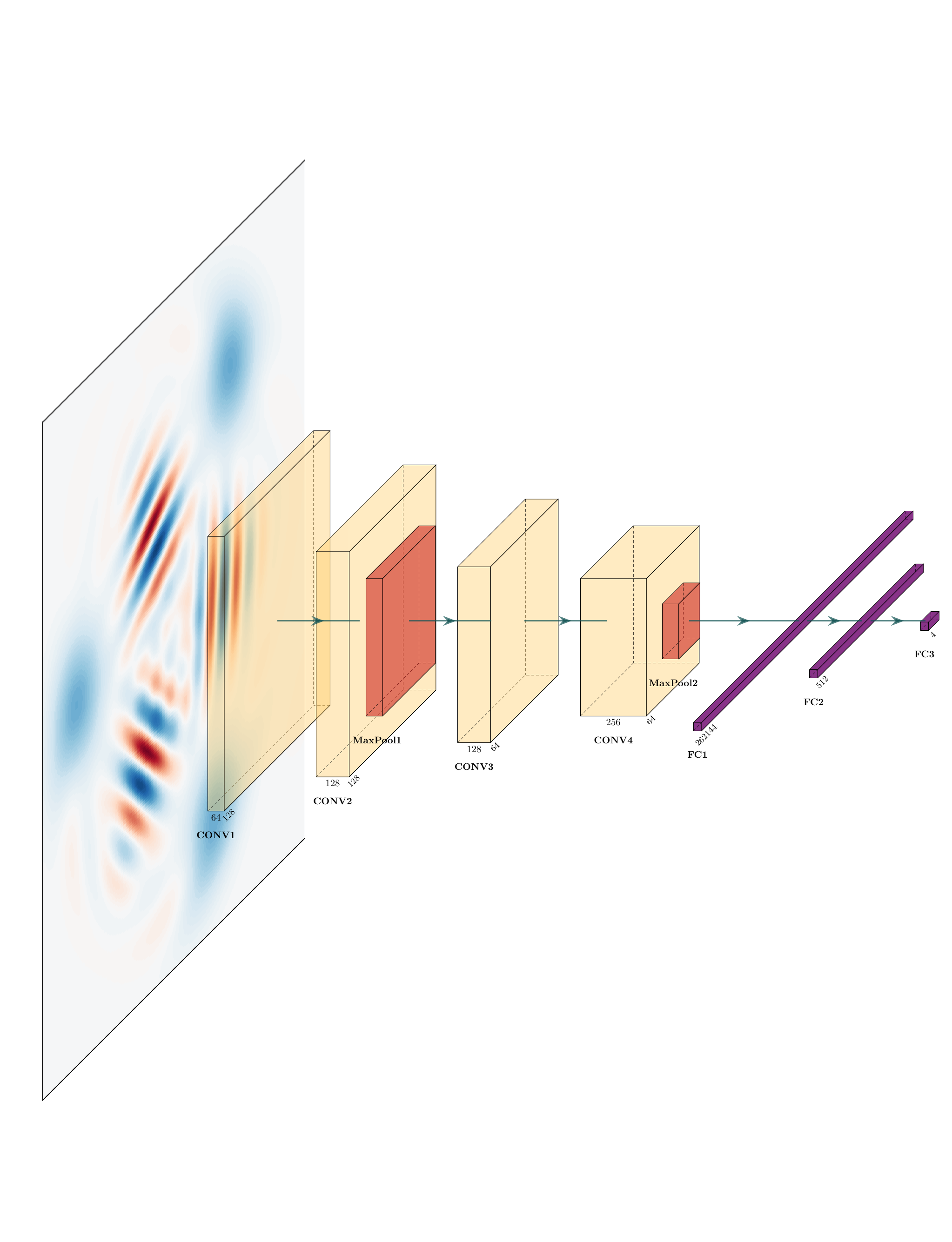}% Here is how to import EPS art
\caption{\label{fig:2} Schematic diagram of LeNet's network structure. The model comprises four convolutional layers, two max pooling layers, two fully connected layers, and one output layer. Yellow squares denote convolutional layers, red squares denote max pooling layers, and purple squares denote fully connected layers and the output layer.}
\end{figure}

The network model we first use is the classic LeNet network structure, as shown in Fig.~\ref{fig:2}. It contains four convolution layers, two pooling layers, two full connection layers and one output layer. The convolution kernel of all convolution layers is 3*3, the step size is 1, the pooling method is maximum pooling, and the activation function is ReLU. The formula for calculating the output image through the convolution layer is

\begin{equation}
\begin{split}
W_{output}=\frac{W_{input}-W_{kernel}+2padding}{stride}+1,
\end{split}
\end{equation}

Where padding is the number of padding outside the image during convolution, and $W_{kernel}$ is the size of convolution kernel. Through this formula, we can calculate the output of each convolution layer to select the appropriate convolution kernel and step size. After every two convolution layers, we add a maximum pool layer to reduce the amount of data in the characteristic graph and reduce the computational complexity. Through the feature extraction of four convolution layers, the feature quantity of the final full connection layer reaches 262144. After two full connection layers, an activation function and a dropout layer, four classifications are finally obtained.

Dropout layer can randomly delete some cells to reduce the complex co adaptation relationship between neurons. Therefore, a hidden layer neuron cannot rely on other specific neurons to correct its errors, forcing the network to learn more robust features. The relu activation function has more efficient gradient descent and back propagation, and avoids gradient explosion and gradient disappearance to a certain extent. For the input vector x from the upper layer neural network entering the neuron, the neuron using the relu function will output

\begin{equation}
max(0,w^Tx+b),
\end{equation}

The loss function we selected is the cross entropy loss function, which can be expressed as

\begin{equation}
H(p,q)=-\sum_{i=1}^np(x_i)\log(q(x_i)),
\end{equation}

Where $p(x_i)$ is the real mark of the ith sample, and $q(x_i)$ is the model prediction probability of the ith sample. Optimizer we choose the adam optimizer, which adjusts the learning rate of each parameter by calculating the first-order moment estimation and the second-order moment estimation of the gradient, so as to achieve more efficient network training.

\subsection{ResNet}

The second network structure we use is the residual network model \cite{al2021automatic,odusami2021analysis,liu2021magnetic,yu2019abnormality}, as shown in Table~\ref{tab:1}. We show the category, output shape and data volume of each layer in the neural network used. In the back-propagation process of neural network, some weight gradients are close to or become zero, resulting in that these weights are hardly updated, which hinders the network training. Residual blocks are the basic building blocks in deep residual networks. By using residual blocks, ResNet can effectively solve the gradient vanishing problem and train very deep networks.

In the traditional convolutional neural network, each convolution layer attempts to learn the mapping between input and output, and the residual block mainly learns the residual mapping between input and output, that is

\begin{equation}
F(x)=H(x)-x,
\end{equation}

Where F(x) is the residual function, H(x) is the objective mapping function, and x is the input. Then add the residual function F(x) to the input x to get the final mapping function

\begin{equation}
H^\prime(x)=F(x)+x,
\end{equation}

\begin{figure}
\includegraphics[width=0.48\textwidth]{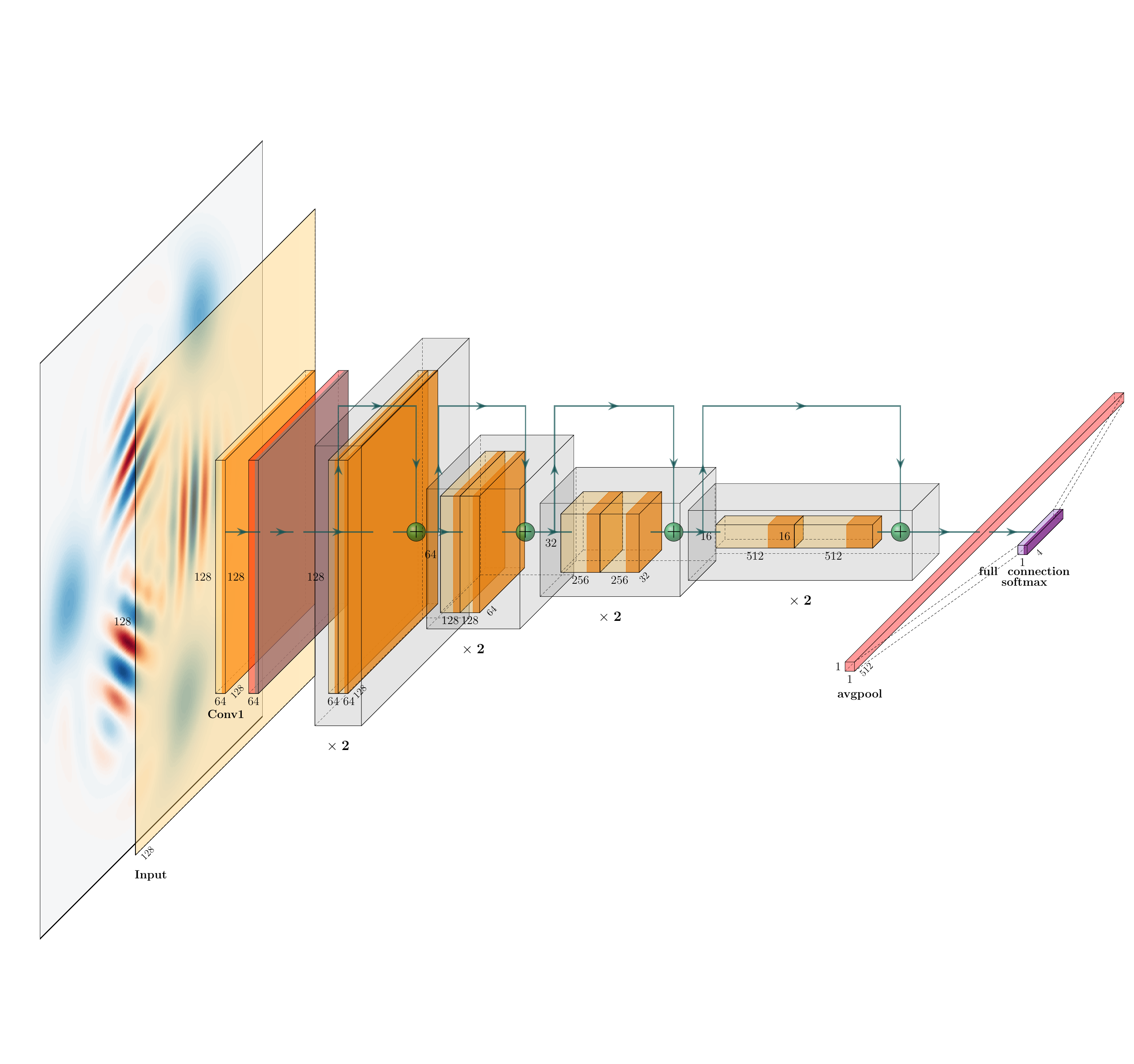}% Here is how to import EPS art
\caption{\label{fig:3} Schematic diagram of ResNet's network structure. It comprises one initial convolution layer, four residual blocks, one average pooling, one full connection layer, and one output layer. Each residual block comprises two convolution layers and two batch normalization layers (not depicted in the figure), with each training cycle involving two residual blocks. Yellow squares denote convolution layers, red squares denote batch normalization layers, pink squares denote average pooling layers, and purple squares denote both the full connection layer and the output layer.}
\end{figure}

Fig.~\ref{fig:3} shows the schematic diagram of the network structure of ResNet, including the initial convolution layer, four residual block groups, an average pooling layer and a full connection layer. The initial convolution layer is used for preliminary feature extraction and spatial down sampling of the input image to a certain extent. It can reduce the spatial dimensions that subsequent layers need to deal with, thus reducing the computational complexity. The convolution kernel of the initial convolution layer we used is 3*3, the step size is 1, and the filling number is 1. The basic features of the image are captured to lay the foundation for subsequent feature extraction.

\begin{table}
\caption{\label{tab:1}The neural network layer of ResNet, the ouput shape and parameter of the output characteristic graph of each layer.}
\begin{tabular}{c c c}
\toprule [1pt]
Layer & Output shape & No. of Parameters \\
\midrule [1pt]
Conv2d(64,3,1) & 128,128,64 & 2368 \\ 
BatchNorm2d & 128,128,64 & 64 \\ 
Conv2d(64,3,1) & 128,128,64 & 36864 \\ 
BatchNorm2d & 128,128,64 & 64 \\
Conv2d(64,3,1) & 128,128,64 & 36864 \\ 
BatchNorm2d & 128,128,64 & 64 \\
Conv2d(64,3,1) & 128,128,64 & 36864 \\ 
BatchNorm2d & 128,128,64 & 64 \\
Conv2d(64,3,1) & 128,128,64 & 36864 \\ 
BatchNorm2d & 128,128,64 & 64 \\
Conv2d(64,3,2) & 64,64,128 & 73728 \\ 
BatchNorm2d & 64,64,128 & 128 \\
Conv2d(128,3,1) & 64,64,128 & 147456 \\ 
BatchNorm2d & 64,64,128 & 128 \\
Conv2d(128,3,1) & 64,64,128 & 147456 \\ 
BatchNorm2d & 64,64,128 & 128 \\
Conv2d(128,3,1) & 64,64,128 & 147456 \\ 
BatchNorm2d & 64,64,128 & 128 \\
Conv2d(128,3,2) & 32,32,256 & 294912 \\ 
BatchNorm2d & 32,32,256 & 256 \\
Conv2d(256,3,1) & 32,32,256 & 589824 \\ 
BatchNorm2d & 32,32,256 & 256 \\
Conv2d(256,3,1) & 32,32,256 & 589824 \\ 
BatchNorm2d & 32,32,256 & 256 \\
Conv2d(256,3,1) & 32,32,256 & 589824 \\ 
BatchNorm2d & 32,32,256 & 256 \\
Conv2d(256,3,2) & 16,16,512 & 1179648 \\ 
BatchNorm2d & 16,16,512 & 512 \\
Conv2d(512,3,1) & 16,16,512 & 2359296 \\ 
BatchNorm2d & 16,16,512 & 512 \\
Conv2d(512,3,1) & 16,16,512 & 2359296 \\ 
BatchNorm2d & 16,16,512 & 512 \\
Conv2d(512,3,1) & 16,16,512 & 2359296 \\ 
BatchNorm2d & 16,16,512 & 512 \\
AdaptiveAvgPool2d & 1,1,512 & 67108864 \\ 
Linear & 1,1,4 & 2048 \\
\bottomrule [1pt]
\end{tabular}
\end{table}

After the initial convolution layer, we use four groups of residual blocks, each of which includes two residual blocks, with 64,128,256 and 512 output channels respectively, and a total of 16 convolution layers. The first residual block in each group reduces the size of the feature map and increases the number of output channels to ensure the computational efficiency of the model, while seizing more hierarchical features. Considering the characteristics of the classification of coherent states and Schrödinger cat states, the residual block we use here is a basic block structure. Through residual linking, each residual block group can learn the complex nonlinear mapping between input and output, extract higher-level features layer by layer, and optimize gradient flow through residual linking, so as to classify images more effectively.

After the residual block group, we use the global average pooling. By reducing each characteristic graph to a single value, we significantly reduce the model parameters and the amount of calculation, improve the generalization ability of the model, and prevent over fitting. The output of the last residual block layer is a 16*16*512 feature map. After passing the average pooling layer, the output becomes a 1*1*512 feature map. At last, we use a full connection layer to classify and regress the 512 data, and get 4 classification numbers.

\section{Training and results}

\begin{figure}
\includegraphics[width=0.48\textwidth]{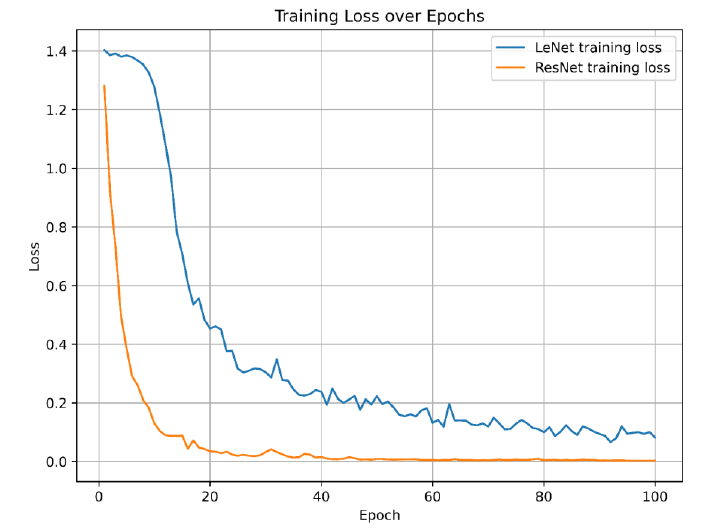}% Here is how to import EPS art
\caption{\label{fig:4} Relationship between loss value and epochs. The figure shows the loss function value curve of LeNet and ResNet on the training set respectively, and the value of epoch is an integer from 0 to 100.}
\end{figure}

The hyperparameters used in training ResNet are shown in Table~\ref{tab:2}, whereas the hyperparameters for LeNet were mentioned earlier. We fed the network model with batches of 20 packets from the dataset for training, iterating through the dataset 100 times, and iteratively performing forward propagation, loss calculation, backward propagation, and parameter updates. By averaging the loss function at each iteration, we can create a graph of loss values against the number of iterations. We obtained the loss values for both LeNet and ResNet, as illustrated in Fig.~\ref{fig:4}.

\begin{table}
\caption{\label{tab:2}Hyperparameters for the ResNet Model.}
\begin{tabular}{c c}
\toprule [1pt]
hyperparameters & Name/Value \\
\midrule [1pt]
Activation function & ReLU \\ 
Rate of learning & $1*10^{-5}$ \\ 
Epochs & 100 \\ 
Batch size & 16 \\
Loss function & Cross Entropy \\ 
Optimizer & Adaptive Moment Estimation(Adam) \\
\bottomrule [1pt]
\end{tabular}
\end{table}

The graph indicates that when classifying quantum states, LeNet's loss value initially decreases rapidly after a relatively stable period, then levels off with fluctuations. This could be due to iterations being constrained by excessively small initialization weights. Overall, LeNet's performance on this classification task is adequate, though it exhibits underfitting.

Regarding the loss curve for ResNet, the loss value decreases rapidly and approaches zero asymptotically, potentially indicating overfitting. By comparison, ResNet converges significantly faster than LeNet and demonstrates stronger generalization capabilities.

\begin{figure}
\includegraphics[width=0.48\textwidth]{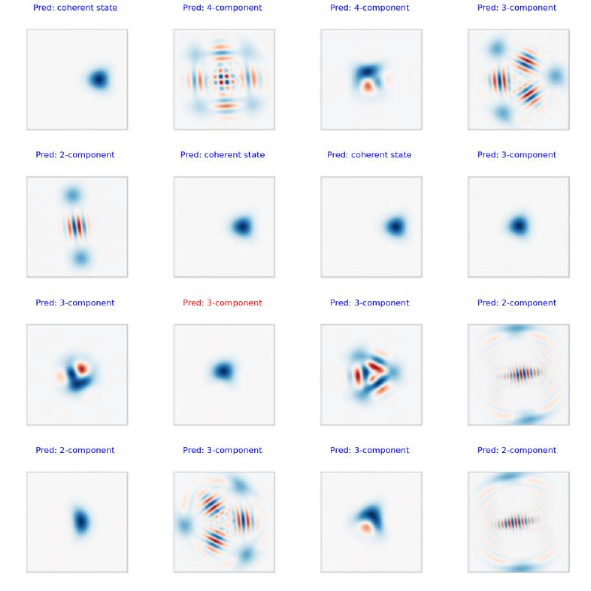}% Here is how to import EPS art
\caption{\label{fig:5} Visualization of the LeNet predictive model. The correct prediction is displayed in blue, and the wrong prediction is displayed in red.}
\end{figure}

Subsequently, we will validate the trained LeNet and ResNet on the test set individually. On 80 test sets, LeNet achieved an accuracy of 97.5$\%$, whereas ResNet reached 100$\%$, indicating that our model can effectively differentiate between coherent states and Schrödinger cat states, addressing the issue mentioned in reference \cite{Ahmed2021} where the network incorrectly identified cat states as coherent states.

We have visualized the incorrect predictions made by the LeNet model, as illustrated in Fig.~\ref{fig:5}. It is evident that LeNet struggled to clearly categorize when the coherent features of the Schrödinger cat state vanished, an issue that is effectively resolved by the ResNet network architecture.

Increasing quantum state data volume and more complex quantum state classification challenges also pose a significant test for ResNet. Building on this foundation, we can explore incorporating various types of Wigner function diagrams and additional quantum states for classification, thereby enhancing the model's generalization capabilities and mitigating the risk of overfitting. Overall, we have introduced a viable approach to the quantum state classification problem: employing ResNet for both training and prediction.

\section{CONCLUSION}

We constructed LeNet and ResNet models, configured with the same loss function and optimizer, and trained them on a dataset of 320 images of coherent and cat states with different $\alpha$. The training process demonstrated that ResNet facilitates more effective gradient descent, avoids overfitting, and more precisely identifies the surface features of coherent and cat states, offering superior performance in quantum state classification. Ultimately, when trained separately on the training set, LeNet and ResNet achieved accuracies of 97.5$\%$ and 100$\%$ respectively. 

This study aims to demonstrate the potential of ResNet in quantum state recognition, highlighting its strengths in handling high-dimensional and complex data, and offering innovative approaches for the future development of quantum information processing and state classification technologies. This advancement will not only enhance the capability to identify Schrödinger cat states but could also spur the exploration of new research directions and application scenarios in quantum computing, communication, and measurement.

\nocite{*}
\bibliography{main}

\end{document}